# High field magneto-transport and magnetization study of Y$_{1-x}$Ca$_x$Ba$_2$Cu$_3$O$_{7-\delta}$ (x = 0.0-0.25)

N. P. Liyanawaduge[1,2,3], Anuj Kumar[2], Rajveer Jha[2], B. S. B Karunarathne[3], and V. P. S. Awana[2*]

[1]*Industrial Technology Institute, P.O Box 363, Baudhaloka Mawatha, Colombo-07, Sri Lanka*

[2]*Quantum Phenomena and Application Division, National Physical Laboratory (CSIR)
Dr. K. S. Krishnan Road, New Delhi-110012, India*

[3]*Postgraduate Institute of Science, University of Peradeniya, Peradeniya (20400), Sri Lanka*

## Abstract

We report *DC* isothermal magnetization, global critical current density ($J_C$), Intra-grain critical current density ($J_C^{intra}$) and resistive upper critical field ($B_{c2}$) of polycrystalline Y$_{1-x}$Ca$_x$Ba$_2$Cu$_3$O$_{7-\delta}$ (YBCO); x = 0.00, 0.05, 0.10, 0.15, 0.20 and 0.25. Ca doping at Y sites in YBCO superconductor results in flux pinning property, which can enhance the critical current density and upper critical fields. $J_C$ and $J_C^{intra}$ are experimentally calculated through *DC* isothermal magnetization measurement (*M-H*), employing the well known Bean's critical model. It is observed that both type of current densities (global and intra) enhance with lower doped (5%) samples and can be accounted in terms of reaching maximum of pinning centers and improved grain couplings nature. The upper critical field ($B_{c2}$) increases with Ca content being consistent with enhanced grain coupling nature of doped samples.

*Corresponding author's E-mail: awana@mail.nplindia.ernet.in

**Tel.** +91-11-45609357; **Fax**. +91-11-45609310.

# 1. Introduction

Following the discovery of high-temperature superconductivity (HTSc) by Bednorz and Muller [1] in metal oxide cuprate, lot of similar materials have been explored with high values of superconducting transition temperature ($T_c$). The discovery of superconductivity in Y-Ba-Cu-O system with transition temperature above 90 K was one of the great achievements in the field of high temperature superconductivity [2]. Although, these materials possess high transition temperature, the practical applications of these materials are yet scant due to relatively inferior values of some superconducting parameters such as critical current density ($J_C$) and upper critical field, ($B_{C2}$) [3]. Therefore, lot of efforts had been put towards discovery of superconducting materials with higher critical parameters. The major limiting factors in case of practical applications of most HTSc systems are inter-grain weak links and poor flux pinning capability. From this point of view, the investigation on the Ca doped YBCO (YBa$_2$Cu$_3$O$_{7-\delta}$) system is of great interest. Calcium ion (Ca$^{2+}$) preferentially substitutes for Yttrium ion (Y$^{3+}$) because of it's ionic radius being closer to that of Yttrium. We have shown in our previous article [4] that Ca doping results in flux pinning behavior and also enhances the inter-granular couplings. Recently R. F. Klie et. al. [5] reported that Ca doping at Y site of YBCO improves inter-grain $J_C$ by controlling the generation of oxygen vacancies at grain boundaries.

The investigation of resistivity transition under an applied magnetic field is an important tool for exploring the percolation nature between grains and grain boundary resistance. Basically the resistive transition occurs in two steps and be understand as the steep transition near the onset part associated superconductivity in individual grains and long transition tail due to the couplings regimes between grains or connective nature of grains [6-9]. Plotting the temperature derivative of resistivity data gives more insight to this two steps structure of resistivity transition. It is essential to look at the temperature derivative of the resistivity in order to give a proper description of the superconducting transition [6]. Correspondingly, temperature derivative of resistivity data gives narrow intense maxima approximately at $T_C$ and broad peak at low temperatures [6-9]. In addition to that, the two peak behavior can be seen even in single crystals of YBCO with high oxygen contents, revealed in heat capacity and resistivity measurements [10]. This is consistent with multi-component (anisotropic) nature of order parameter or structural phase separation [7, 10- 11]. On the other hand the resistive upper critical field, derives from resistivity measurements, gives information on superconducting properties and allows calculating microscopic parameters.

In this paper we report structural/micro-structural, high field magneto-transport and magnetic measurements of $Y_{1-x}Ca_xBa_2Cu_3O_{7-\delta}$; x = 0.00, 0.05, 0.10, 0.15, 0.20 and 0.25 and describe the persisting flux pinning behavior and improved grain-connectivity of doped samples. Subsequently the enhancement in critical current densities and upper critical field are discussed in detail in terms of above enhanced pinning and improved grain-connectivity of doped samples.

## 2. Experimental

All samples were synthesized through the conventional solid state reaction route. High purity powders of $Y_2O_3$, $CaCO_3$, $BaCO_3$ and $CuO$ and (99.99%) with the exact stoichiometric ratio were mixed. After initial grinding, calcinations was done at $850^oC$ in air for 36 h and cooled to the room temperature over a span for another 12 h. Subsequently, calcinations were done at $870^oC$, $890^oC$ and $910^oC$ temperatures for same duration with intermediate grinding. In each calcinations cycle cooling was done slowly and samples were re-ground well before the next cycle. After final calcinations all samples were pressed into a rectangular pellet form and sintered at $925^oC$ for 48 h in air. Finally samples were annealed with flowing oxygen at $750^oC$ for 12 h, 600°C for 24 h and 450°C for 24 h and subsequently cooled to room temperature in 6 h. All samples were characterized by the X-ray powder diffraction technique using Rigaku X-ray diffractometer (CuK$_\alpha$ radiation, λ = 1.54 Å). Rietveld analysis of all samples was performed using *Fullprof* program [12]. The DC magnetic susceptibility (χ-*T*), isothermal magnetization (*M-H*) and resistivity measurements with and without magnetic field *RT (H)* were performed on Physical Properties Measurement System (PPMS-14T, Quantum Design-USA). The Scanning Electron Micrograph (*SEM*) images and corresponding *EDX* measurements of freshly fractured surfaces of samples were taken using *ZEISS* EVO MA-10 Scanning Electron Microscope.

## 3. Results and Discussion

The Rietveld fitted *XRD* patterns of all the studied samples are shown in Fig. 1 (a-b). The structural analysis was performed using the Rietveld refinements with help of *FullProf* software. Rietveld analysis confirmed the single phase formation of all samples in space group *pmmm*. The lattice parameters and other fitted parameters are given in Table 1. After examining the lattice parameters it is observed that samples belong to oxygen rich regime at lower doped levels [13] and orthorhombicity decreases with increasing Ca concentration [14, 15]. The *c* parameter increases with Ca doping, indicating Ca substitute in eight fold coordination number with ionic radius 1.12 Å, in similar coordination number to the existing coordination number of Y with ionic radius 1.0 Å. The Scanning Electron Micrograph (*SEM*) images of freshly fractured surfaces of samples with x = 0.00, 0.05, 0.015 and 0.20 and electron dispersive X-Ray (*EDX*) spectrums of samples x = 0.00 and x = 0.20 are given in Fig.2. It is observed from the *SEM* images that the pristine sample is having better surface texture, high porosity and lower density. The grain connectivity has increased with Ca doping. The circular grains are contained in pure sample, while the doped samples contain the elongated grains and this is a clear indication for changing of grain morphology with Ca doping. The grain size has initially increased at small doping (x = 0.05) concentrations and then slightly decreased at further increase of Ca content. $Ca^{2+}$ not only goes to the grain but also to grain boundaries due to the highly segregation ability of Ca [16]. Therefore some $Ca^{2+}$ segregated to surface of grains and thus controlled the grain growth [17] at higher level doped samples. The sample with x = 0.05 shows good crystallization with elongated larger grains. The atomic composition of the samples was measured by *EDX* measurement and it is confirmed that samples are approximately in intended atomic ratios as they were mixed and do not contain foreign particles.

The samples were cooled in zero magnetic fields down to the lowest accessible temperature (2 K). After temperature stabilization a DC magnetic field of 10 Oe is applied and the magnetization recorded as a temperature raised (ZFC curve) up to 100 K. Then the measurement continued while the temperature was again decreased back to 2 K (FC curve), keeping the same DC magnetic field. Fig. 3 presents the variation of dimensionless DC volume susceptibility curves in both Zero Field Cooled (ZFC) and Field Cooled (ZFC) situations. It is remarkable that the separation of FC and ZFC signals increases with Ca doping. The increase in separation of FC and ZFC signals or in other words reduction of Meissner fraction (ratio of field cooled to zero fielded cooled magnetization) is a clear indication for flux pinning of these samples. Therefore, an enhancement in critical current densities and critical fields can be expected in doped samples.

Magnetization measurement is very informative to understand the critical properties of type-II high $T_c$ superconductors. The isothermal magnetization loops (*M-H)* up to 90 kOe at 5 K, 20 K, 50 K and 70 K temperatures of the samples with x = 0.00, 0.05 and 0.15 are given in Fig. 4. The applied magnetic field is parallel to the long axis of the measured samples. It is obvious that diamagnetism and remnant magnetization increases in x = 0.05 than pure sample and subsequently decrease in higher doped samples. Moreover the sample with x= 0.05 shows higher magnetization irreversibility ($|\Delta m|$) at all values of applied fields than pure sample. However this is decreased in sample with x = 0.15 at all values of applied field. Insets of Fig. 4 depicts the forth quadrant of magnetization loops which indicates the lower critical field, ($B_{C1}$). It is observed that $B_{C1}$ is maximum for x = 0.05. Clearly, the magnetization behavior has been improved for lower doped samples (x = 0.05) in comparison to both pristine and higher (x > 0.05) Ca doped samples.

Potential applications of these superconducting materials require reasonably higher super current densities. Therefore, the magnetization data were interpreted to yield global critical current values ($J_C$) up to 90 kOe at 5 K temperature, explained in Bean's model [18] given by

$$J_C = \frac{20\Delta m}{Va(1 - a/3b)}$$

where, $\Delta m$ is measured in emu, $V$ is in cm$^3$, $a$ and $b$ ($a < b$) are in cm and $J_C$ is in A/cm$^2$. The evaluated magnetic field dependency of $J_C$ is shown in Fig. 5. It can be seen that clear enhancement of current density in x = 0.05 sample than parent sample but decrease in higher doped samples for all values of field. It is known that both intra-grain currents and inter-grain currents contribute to the magnetization of the bulk sample. Also, intra-grain and inter-grain critical current densities are determined by collective flux creep inside the grains and the maximum Josephson current across the grain boundaries (flux flow) respectively [19, 20]. Therefore this enhancement of global $J_C$ of doped sample can be consequence of both pinning property and enhancement of Josephson current across grain boundaries by improvement of grain couplings as revealed in references [4, 5]. H. Huhtinen et.al [21] accounted of this high $J_C$ behavior of Ca doped samples by higher carrier concentrations at grain boundaries with presence of pinning nature by normal nano dots created by oxygen vacancies in Cu-$O_2$ planes. Song et.al [22] explained this positive effect of Ca doping as enhancement of grain boundary critical current by expansion of dislocation cores due to segregated $Ca^{2+}$ at grain boundaries.   Mean while this enhancement of super current in Ca doped YBCO system has been described in terms of promoting of carrier density by controlling the generation of oxygen vacancies at grain boundaries which can enhance the grain boundary current [5, 23-25].

Assuming all the current loops are restricted within the grains with an average radius R, the intra-grain current ($J_C^{intra}$) was calculated by

$$J_C^{intra} = \frac{30\Delta M}{R}$$

The average grain size of studied samples was calculated by standardized linear intercept (LI) method on *SEM* images and found to be order of 2-4 µm. The calculated intra-grain $J_C$ values were plotted in right inset of Fig. 5. The sample with x = 0.05 shows maximum intra-grain $J_C$. Because of intra-grain regions are described by collective flux creep model [19, 20], this enhancement is attributed in terms of flux pinning nature developed in doped samples. The $J_C^{intra}$ correlates well with the condensational pinning energy [26]. The pinning force, $F_P$ of the samples was calculated by $\mu_0 Jc^{intra}H$ and plotted in left inset of Fig. 5.

Even though DC susceptibility data show that pinning centers are developed with x, the *M (H)* data and subsequent current densities have enhanced only in lightly doped samples and degraded at higher doping. In terms of grain size variation, $J_C^{intra}$ should decrease first and then increase at higher doped samples. It is known that Ca substitution enhance the inter-grain current ($J_C^{inter}$) by improving inter-grain couplings. These factors suggest that current densities should enhance further with Ca. But according to the Fig. 5 the current densities degrades for higher level doped samples. Therefore it seems that some factor, which limits the current densities, affects the highly doped samples. Effective pinning, from defects being induced by the stress-field of lattice mismatch [27] could be one such possibility, which can account for this behavior of critical currents. At lower concentrations (around 5%) the density of defects corresponding to stress-field may reach maximum. In addition to that careful examination of SEM images indicate that several partially melted spots are seen in highly Ca doped samples (x = 0.15 and x = 0.20). This is possible because the melting temperature of these polycrystalline samples becomes lower with increase in the Ca concentration. This may also hinder the effective pinning and act as a limiting factor for critical current densities. Also with the decrease of grain size, the number of grain boundaries is increased, which could limit the passage of super currents. This may also act as a limiting factor for global current density [28-29].

In order to investigate broadening of resistivity curves in Ca doped YBCO system, we measured the resistivity under magnetic field up to 13 T, applied perpendicular to the current and flat surface of the samples. The corresponding normalize $\rho_{nor}$ curves are depicted in Fig. 6 for $Y_{1-x}Ca_xBa_2Cu_3O_{7-\delta;}$ x = 0.00, 0.10 and 0.25. These curves show basically two transitions in agreement to early studies [6-9, 30]. It is well known that long-range superconducting state with zero resistance is achieved by means of a percolation like process that overcomes the weak links between grains [30]. The broadening of tail part is consistent with disturbances on percolation path between grains due to poor grain alignments cause by applied field together with anisotropic nature [31, 32]. Samples with x = 0.00 and x = 0.10 show larger shift towards the lower temperatures at the tail part of transitions. On the other hand x = 0.25 shows comparatively less shift at the tail part of the superconducting transition. The reduction in the shift of tail part indicates towards better grain couplings in Ca doped samples, which is in agreement with our *SEM* observations and some early reports [4, 5].

For further investigation of this effect we explored the temperature derivative of normalized resistivity, which is given in Fig. 7. In zero applied fields, the grains maintain good percolation path between them and the transition temperature range corresponds to bulk grains and grain boundaries lies very closely. This gives sharp transition in resistivity (ρ-T) curves. Accordingly one single peak appears in derivative curves in Fig. 7 at zero applied fields. The two peaks correspond to the two regimes separate from each other at different rates as increasing the field. The peak belongs to granular network is shifting towards low temperatures more rapidly than the bulk peak. Simultaneously rapid broadening of coupling peak is visible as in agreement to ρ-T curves. The bulk peak also diminishes with increase of applied field as flux penetrates into the individual grains. But shift and broadening of coupling peak is decreasing with increasing Ca content and become minimum in x = 0.25 sample. The analysis of $\rho_{nor}$ curves and their derivatives once again prove that Ca doping to the YBCO system improves the grain couplings.

Temperature dependency of resistive upper critical field, $B_{c2}$ *(T)*, using midpoint data criteria, where the resistivity is half of its normal state value, is shown in inset of Fig. 8. It is seen that all the samples show concave curvature (upward curvature) near the $T_c$ and linear region afterward. This is in agreement with some early studies [31-33]. It is also noticed that this upward curvature is increasing monotonically from pristine sample to x = 0.25 sample. Werthamer, Helfand, and Hohenberg (WHH) theory gave a solution for linearized Gor'kov equations for $H_{c2}$ for bulk weakly coupled type-II superconductor, including effects of Pauli spin paramagnetism and spin-orbit scattering [34]. Here we use simplified *WHH* equation to estimate $B_{c2}(T)$ without spin paramagnetism and spin-orbit interaction, and given by

$$ln \frac{1}{t} = \psi\left(\frac{1}{2} + \frac{\overline{h}}{2t}\right) - \psi\left(\frac{1}{2}\right)$$

where t = T/T$_c$, $\psi$ is the digamma function and $\overline{h}$ is given by

$$\overline{h} = \frac{4H_{c2}}{\pi^2 T_c (-dH_{c2}/dT)_{T=T_c}}$$

Using slope of linear portion of the experimental data and corresponding extrapolated T$_c$ values to $\mu_o H_{c2} = 0$ region, the $\mu_o H_{c2}(T)$ is fitted up to 0 K using simplified *WHH* model and is given in Fig. 8. The calculated $B_{C2}(0)$ of pristine sample is around 66 Tesla and increases with increasing Ca content of the system to a maximum of above 110 Tesla for x = 0.25. Both $(dB_{c2}/dT)_{Tc}$ and $B_{c2}(0)$ improves with doping of Ca at Y site in YBCO system for lower Ca content due to better grains coupling. On the other hand pinning becomes weaker at higher concentrations of Ca due to segregation of Ca at grain boundaries.

Several studies have been made earlier on single crystals and epitaxial thin films of YBCO and $B_{C2}(0)$ is calculated [35-41]. Due to the anisotropy in coherence length along the parallel and perpendicular directions to the Cu-O$_2$ planes, large anisotropy has been observed for $B_{C2}(0)$ when field is applied parallel and perpendicular to Cu-O$_2$ planes [35-41]. For the single crystals, these values were measured by explosive flux compression technique and were 120 T and 250 T along the parallel and perpendicular directions to the Cu-O$_2$ planes respectively [35]. For the thin films these values were reported as 120 T [41] along the perpendicular to the Cu-O$_2$ planes and more than 240 T in Cu-O$_2$ planes [37, 39]. The findings reported in this study (66 T for pristine YBCO and 120 T for *x* = 0.25 composition) seems reasonable in bulk polycrystalline samples.

## 4. Conclusions

In summary, we presented detailed magnetization and magneto-transport of the $Y_{1-x}Ca_xBa_2Cu_3O_{7-\delta}$ polycrystalline samples. It is found that global and intra-grain critical current densities ($J_C$ and $J_C^{intra}$) enhance with limited doping of Ca (approximately 5%). This enhancement is accounted in terms of reaching maximum of effective pinning centers and improved grain couplings nature. Also an improvement in upper critical field with increasing of Ca level is observed, which is consistent with enhanced grain coupling nature. At high concentrations of Ca, these current densities are low due to partial melting of samples segregation of Ca at grain boundaries. Our results demonstrated that limited doping of Ca at Y site improved profoundly the superconducting performance of Y-Ba-Cu-O, without compromising much on its superconducting transition temperature ($T_c$).

## Acknowledgements

This work is supported by RTFDCS fellowship programme conducted by Center for International Cooperation in Science (CICS). Authors are thankful to Prof. R.C Budhani, Director NPL and to Dr. (Mrs.) Ganga Radhakrishnan, Director CICS for their encouragement. One of the authors Mr. Anuj Kumar would like to thank CSIR for providing financial support through Senior Research Fellowship (SRF) during the work.

# Figure Captions and Table

**Figure 1:** Rietveld fitted *XRD* pattern of $Y_{1-x}Ca_xBa_2Cu_3O_{7-\delta}$, x: 0.00, 0.05, 0.10, 0.15, 0.20 and 0.25 oxygen annealed samples.

**Figure 2:** *SEM* images of $Y_{1-x}Ca_xBa_2Cu_3O_{7-\delta}$; (a) x = 0.00, (b) x = 0.05, (c) = 0.15, (d) = 0.20, at 5,000 magnification and *EDX* spectrum (e) x = 0.00, (f) x = 0.20.

**Figure 3:** Variation of dimensionless DC volume susceptibility with temperature of $Y_{1-x}Ca_xBa_2Cu_3O_{7-\delta}$ samples annealed in oxygen environment.

**Figure 4:** The isothermal magnetization loops (*M-H*) up to 90 kOe applied field at temperatures 5 K, 20 K, 50 K and 70 K of $Y_{1-x}Ca_xBa_2Cu_3O_{7-\delta}$, x: 0.00-(a), 0.05-(b) and 0.15-(c) samples. Insets show the forth quadrant of the *M-H* loops up to 5 kOe indicating the variation of lower critical field, $B_{c1}$.

**Figure 5:** Magnetic field dependency of Global critical current density ($J_C$) at 5 K of $Y_{1-x}Ca_xBa_2Cu_3O_{7-\delta}$, x: 0.00, 0.05 and 0.15 samples, deduced from magnetic hysteresis loops shown in Fig.4. Insets show the field dependency of intra-grain critical current density, $J_c^{intra}$ and pining force, $F_p$.

**Figure 6:** Temperature dependency of normalized resistivity at different dc fields, $\rho_{nor}$ of $Y_{1-x}Ca_xBa_2Cu_3O_{7-\delta}$, x: 0.00-(a), 0.10-(b) and 0.25-(c), samples.

**Figure 7:** Temperature derivative of normalized resistivity of $Y_{1-x}Ca_xBa_2Cu_3O_{7-\delta}$, x: 0.00-(a), 0.10-(b), and 0.25-(c) samples, derived from resistivity data in Fig. 6.

**Figure 8:** Fittings of resistive upper critical field, $B_{c2}(T)$, up to 0 K using simplified WHH theory for $Y_{1-x}Ca_xBa_2Cu_3O_{7-\delta}$, x: 0.00, 0.10 and 0.25 samples. Inset shows the variation of $B_{c2}(T)$ up to 13 T derived from experimental data.

**Table 1:** Rietveld refined parameters of $Y_{1-x}Ca_xBa_2Cu_3O_{7-\delta}$, oxygen annealed samples.

| $Y_{1-x}Ca_xBa_2Cu_3O_{7-\delta}$ | $a$ (Å) | $b$ (Å) | $c$ (Å) | $R_p$ | $R_{wp}$ | $\chi^2$ |
|---|---|---|---|---|---|---|
| x = 0.00 | 3.822(1) | 3.887(1) | 11.666(0) | 4.69 | 5.94 | 2.78 |
| x = 0.05 | 3.823(0) | 3.883(0) | 11.677(3) | 5.32 | 6.87 | 3.95 |
| x = 0.10 | 3.826(1) | 3.881(0) | 11.680(5) | 4.67 | 5.94 | 3.05 |
| x = 0.15 | 3.827(3) | 3.880(1) | 11.684(6) | 4.43 | 5.73 | 2.75 |
| x = 0.20 | 3.831(0) | 3.877(4) | 11.691(2) | 4.91 | 6.31 | 3.05 |
| x = 0.25 | 3.835(1) | 3.870(6) | 11.693(0) | 5.44 | 7.23 | 3.64 |

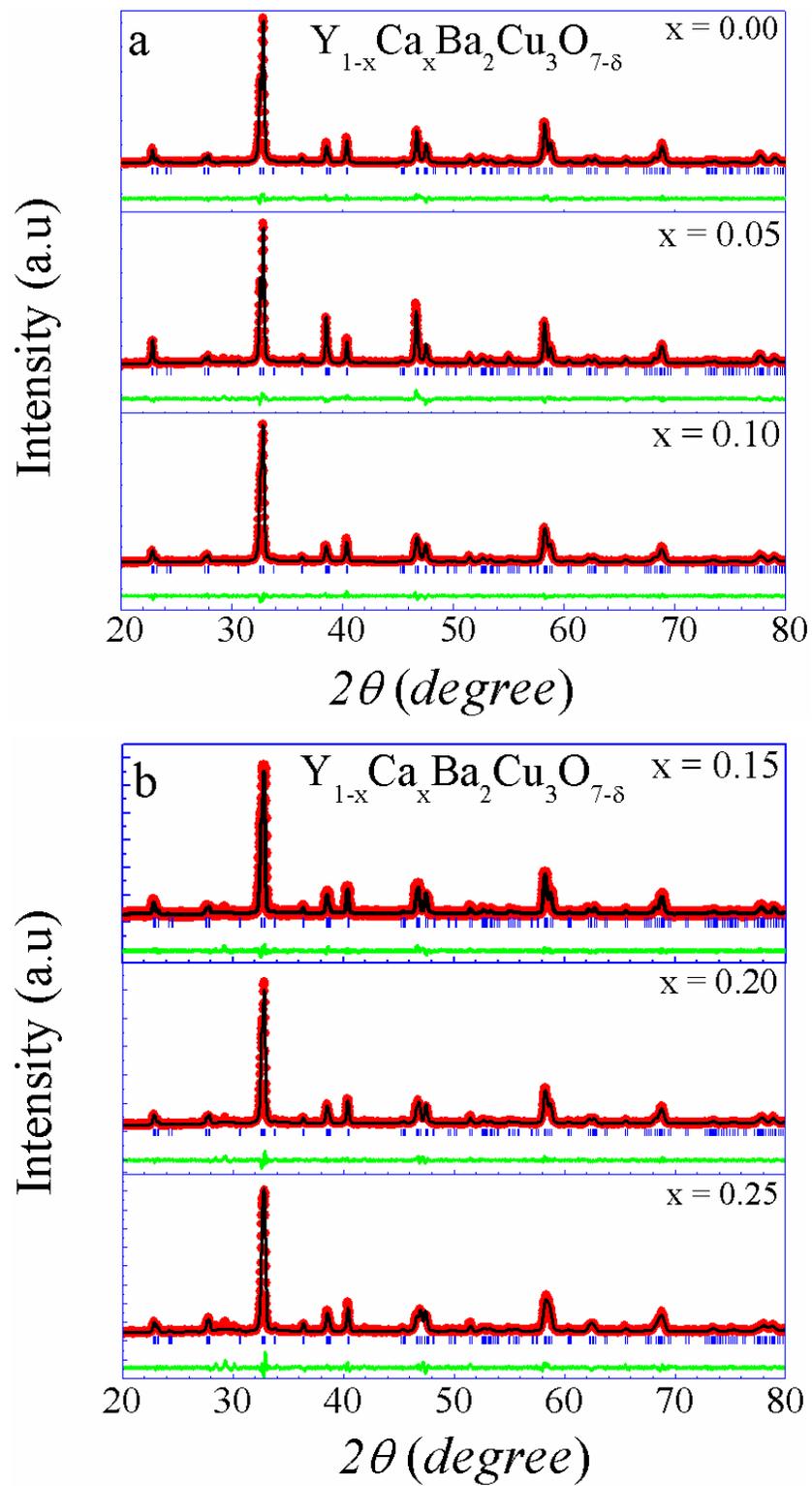

**Figure 2**

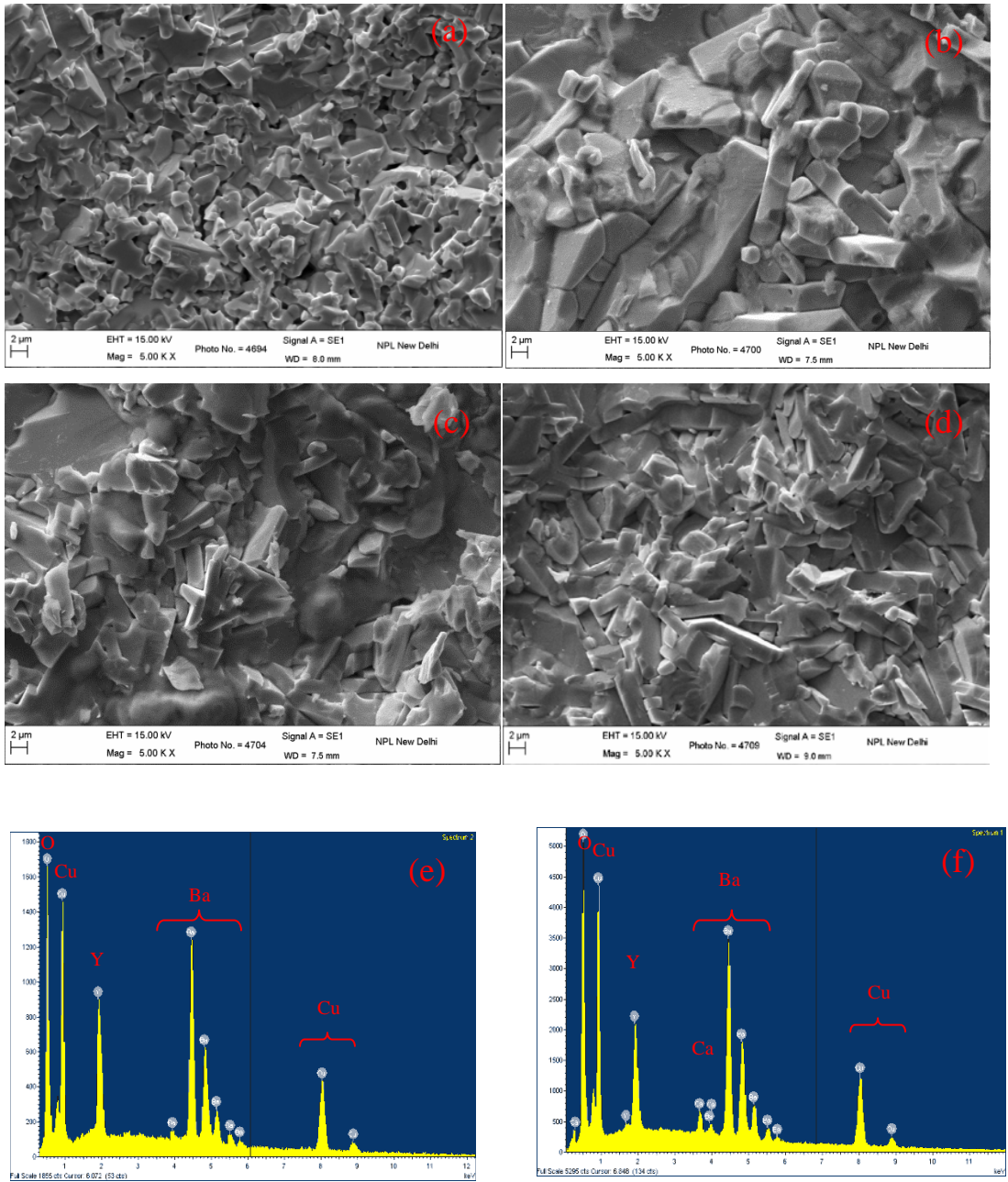

**Figure 3**

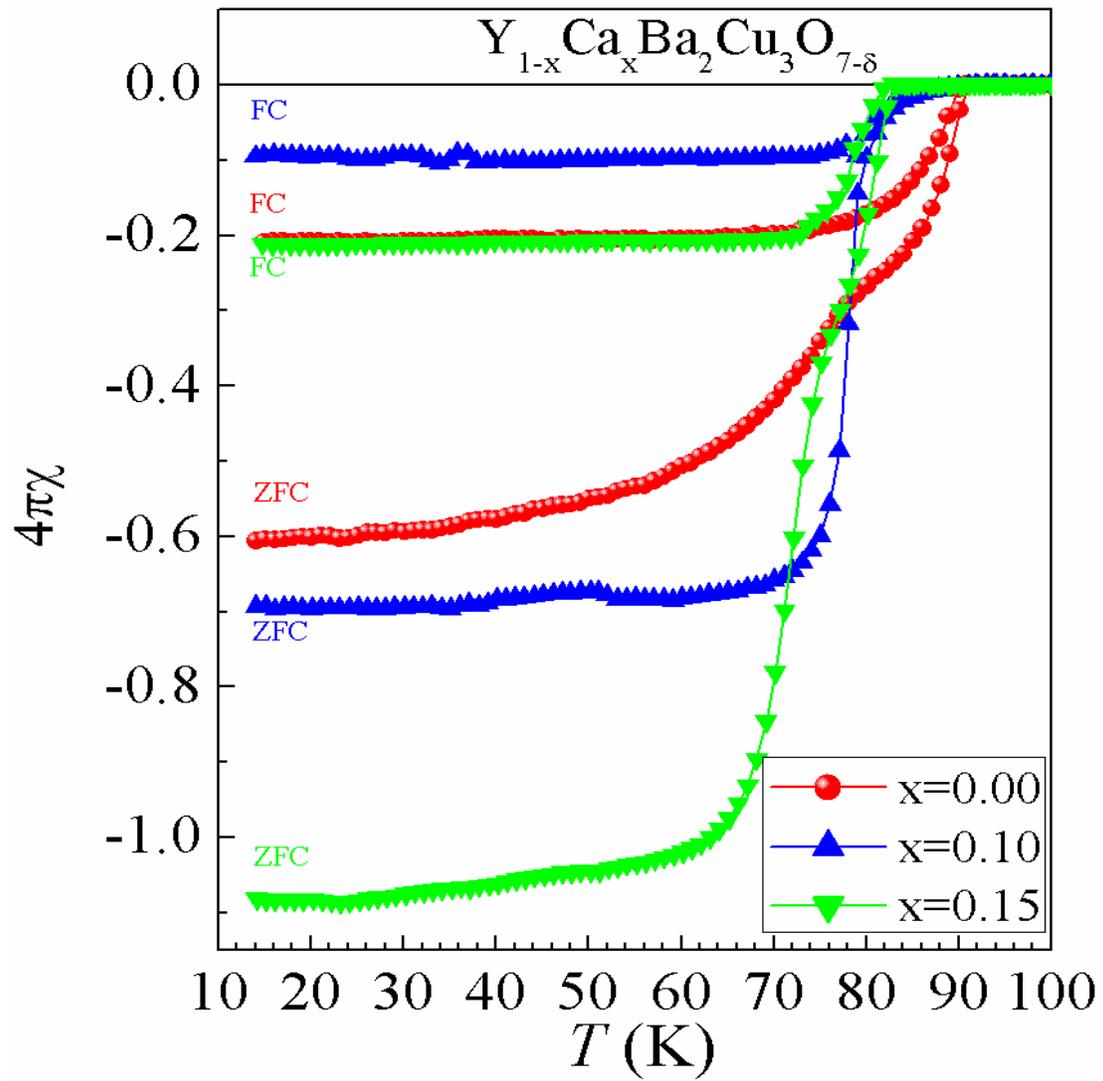

**Figure 4**

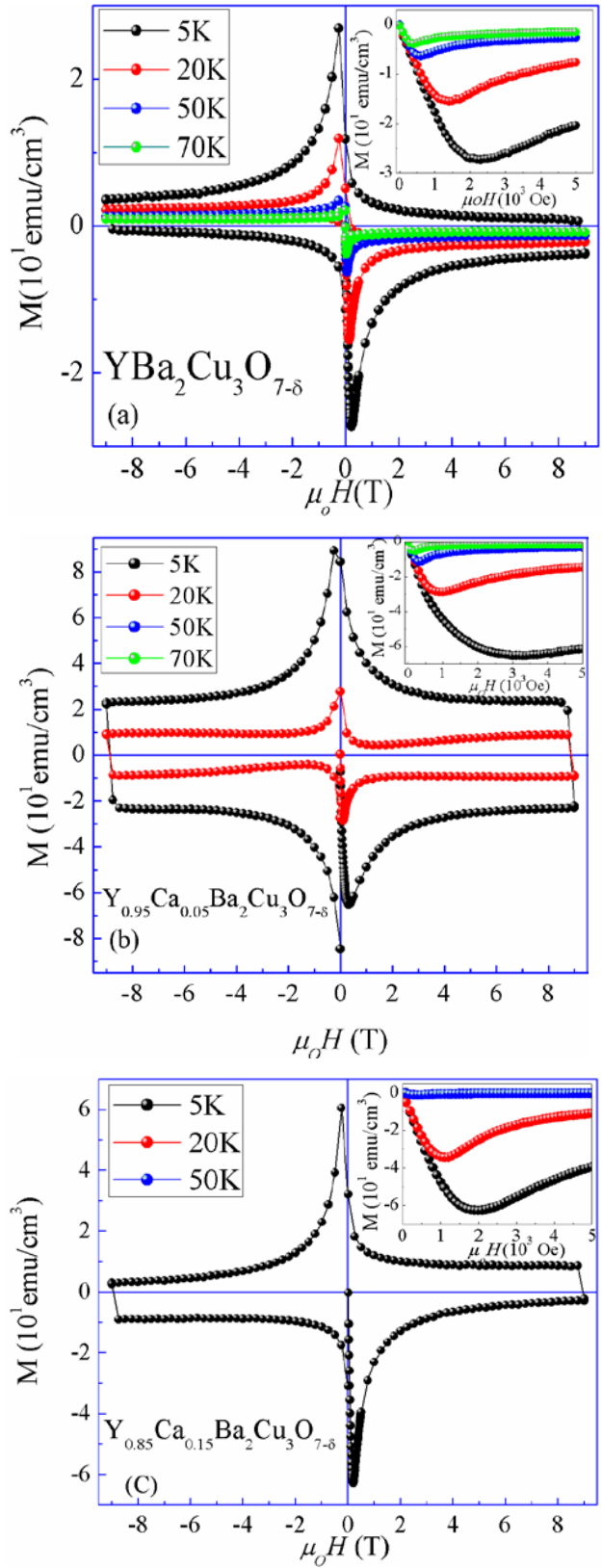

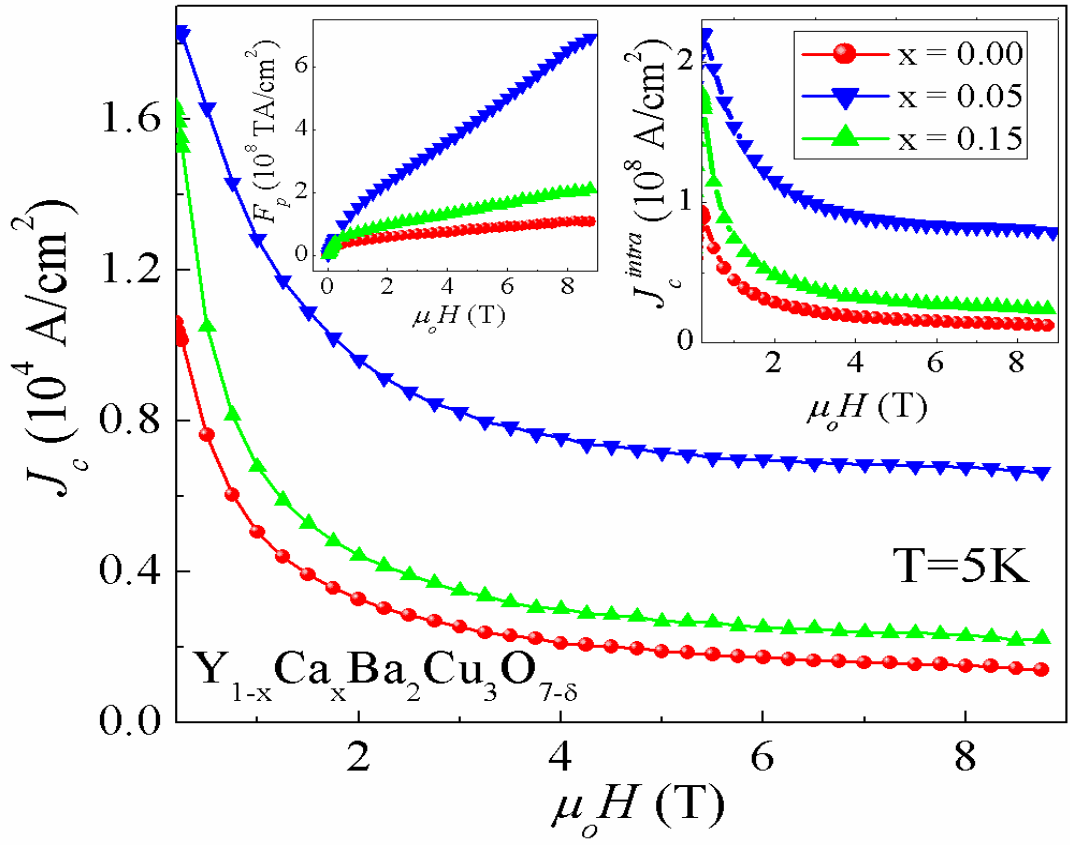

**Figure 6**

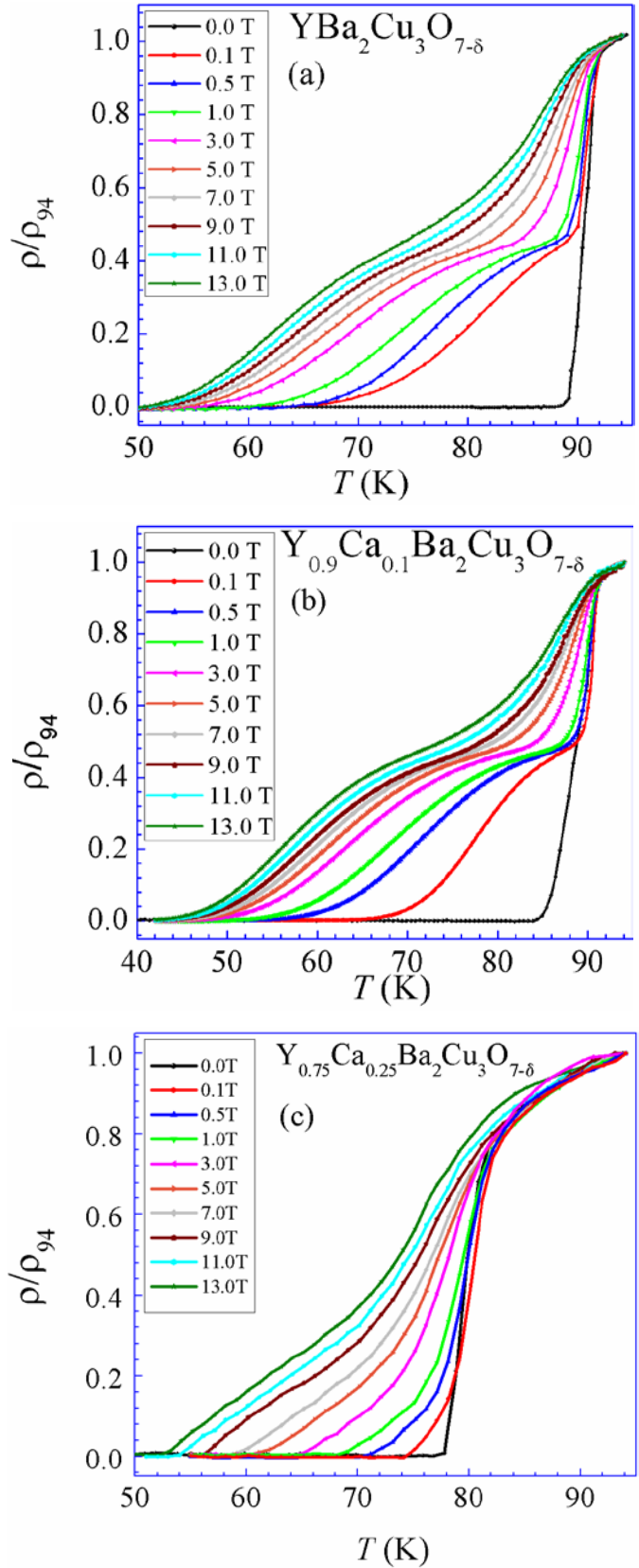

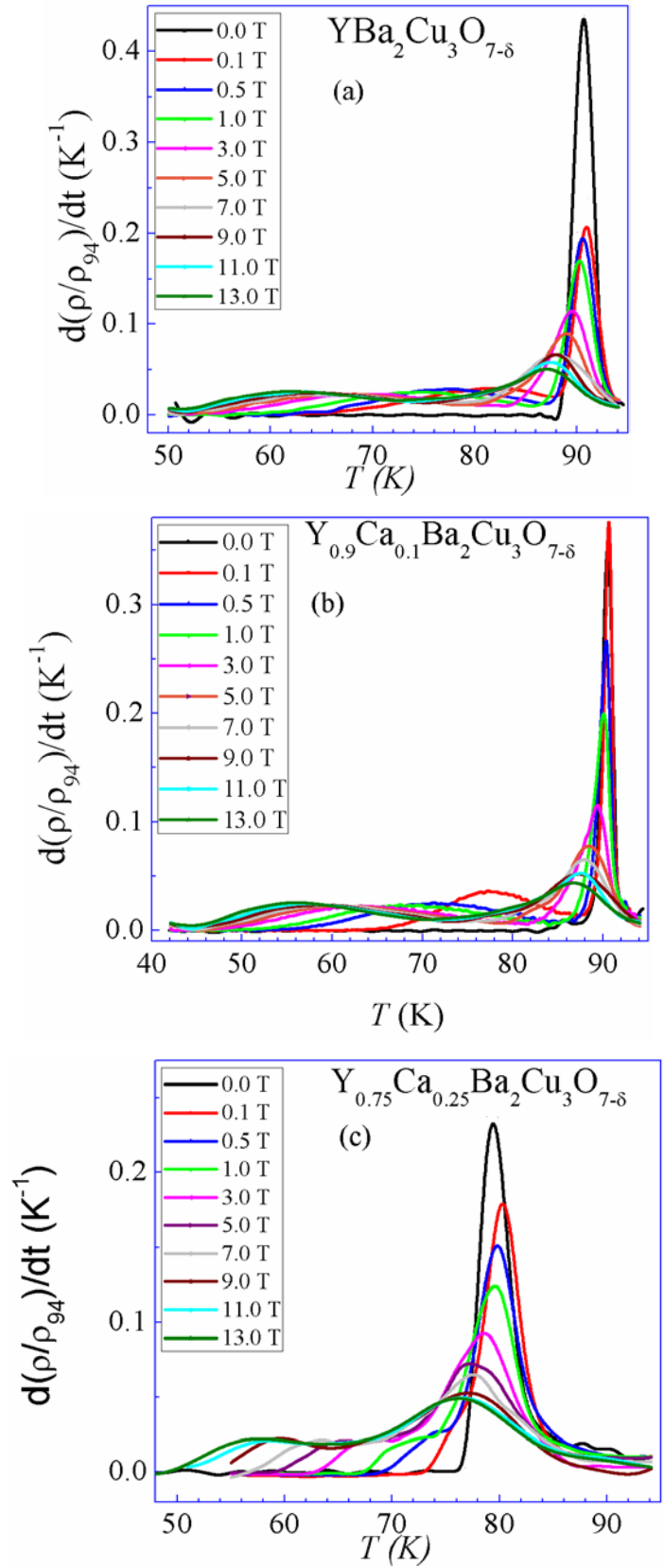

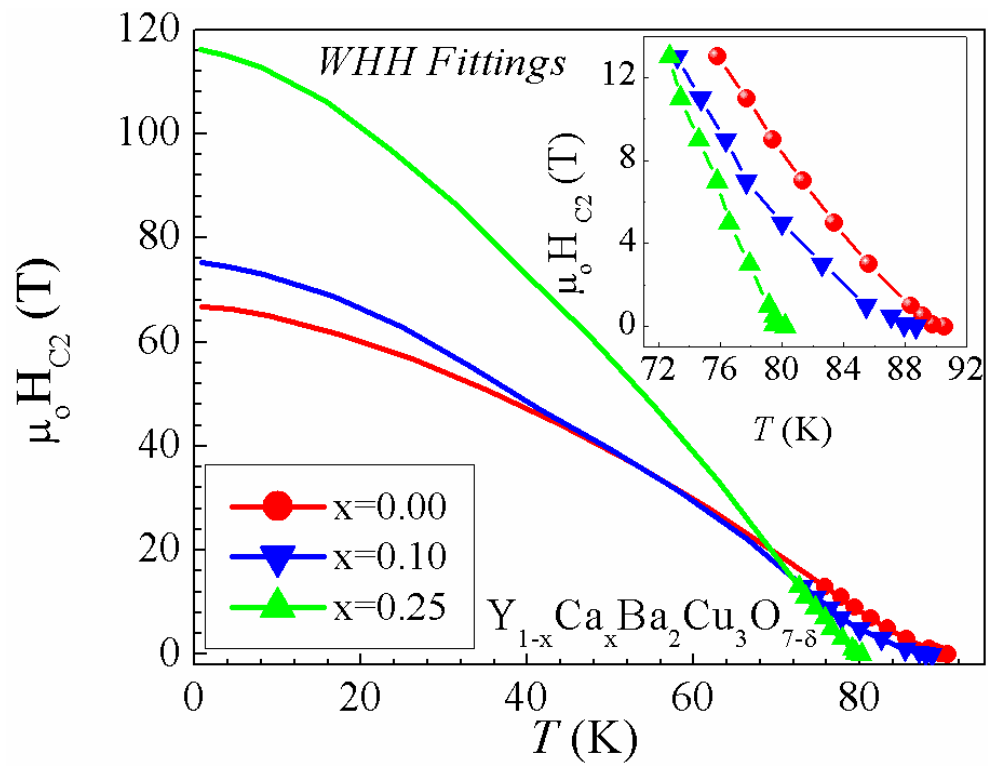